\documentstyle[fortschritte]{article}

\def\beq{\begin{equation}}
\def\eeq{\end{equation}}
\def\6{\langle}
\def\9{\rangle}

\def\0{\otimes}
\def\Tr{{\rm Tr}}
\def\1{\mbox{1\hskip-.25em l}}  

\begin{document}

\title{How the No-Cloning Theorem Got its Name}

\author{Asher Peres\\
Department of Physics,\\ Technion---Israel Institute of
Technology,\\ 32000 Haifa, Israel}

\maketitle
\begin{abstract}
I was the referee who approved the publication of Nick Herbert's {\it
FLASH\/} paper, knowing perfectly well that it was wrong. I explain
why my decision was the correct one, and I briefly review the progress 
to which it led.
\end{abstract}

\bigskip

The no-cloning theorem \cite{wz,dieks} is of fundamental importance
in quantum theory. It asserts that no quantum amplifier can duplicate
accurately two or more nonorthogonal quantum states. A simple
proof requires only a few lines \cite{qt}. Why wasn't that theorem
discovered fifty years earlier? What are the events that actually led
to its discovery and publication?

This is the story of my own personal contribution to the no-cloning
theorem, made public for the first time after more than twenty years.
Early in 1981, the editor of {\it Foundations of Physics\/} asked
me to be a referee for a manuscript by Nick Herbert, with title
``{\it FLASH\/}---A superluminal communicator based upon a new kind of
measurement.'' It was obvious to me that the paper could not be correct,
because it violated the special theory of relativity. However I was sure
this was also obvious to the author. Anyway, nothing in the argument
had any relation to relativity, so that the error had to be elsewhere.

Herbert's apparatus was an idealized laser gain tube which would have
macroscopically distinguishable outputs when the input was a single
arbitrarily polarized photon. Indeed, the word LASER is an acronym for
``Light Amplification by Stimulated Emission of Radiation.'' However,
besides the stimulated emission, there is also {\it spontaneous
emission\/}, which results in noise. Herbert's claim was that the noise
would not prevent identifying the polarization of the incoming photon,
at least statistically. For this, he used the notion of {\it quantum
compounds\/} \cite{compounds} and many properties of laser physics
with which I was not familiar.

I recommended to the editor of {\it Foundations of Physics\/} that this
paper be published \cite{flash}. I wrote that it was obviously wrong, but
I expected that it would elicit considerable interest and that finding
the error would lead to significant progress in our understanding
of physics. Soon afterwards, Wootters and Zurek \cite{wz} and Dieks
\cite{dieks} published, almost simultaneously, their versions of the
no-cloning theorem. The tantalizing title ``A single quantum cannot
be cloned'' was contributed by John Wheeler. How the present paper
got its name is another story \cite{kipling}.

There was another referee, GianCarlo Ghirardi, who recommended to
reject Herbert's paper. His anonymous referee's report contained
an argument which was a special case of the theorem in references
\cite{wz,dieks}. Perhaps Ghirardi thought that his objections were so
obvious that they did not deserve to be published in the form of an
article (he did publish them the following year \cite{gw}). Other
objections were raised by Glauber \cite{glauber}, and then by many
other authors whom I am unable to cite, because of space limitations.

With some hindsight, it is now clear that the no-cloning interdiction
was implicitly used by Stephen Wiesner in his seminal paper {\it
Conjugate Coding\/} which was submitted circa 1970 to IEEE Transactions
on Information Theory, and promptly rejected because it was written in
a jargon incomprehensible to computer scientists (this actually was a
paper about physics, but it had been submitted to a computer science
journal). Wiesner's work was finally published in its original form in
1983 \cite{wiesner} in the newsletter of ACM SIGACT (Association
for Computing Machinery, Special Interest Group in Algorithms and
Computation Theory). Another early article, {\it Unforgeable Subway
Tokens\/} \cite{tokens}, also tacitly assumes that exact duplication
of a quantum state is impossible. As it often happens in science,
these things were well known to those who know things well.

Anyway, the no-cloning theorem was no justification for rejecting
Herbert's paper. His manuscript mentioned exact duplication as a
theoretical ideal, but it was clear that what was actually produced
by the laser gain tube was a messy state, where many different outputs
were entangled with the laser final state.

The no-cloning theorem and its practical implications for quantum
cryptography led to an enormous wave of interest, as I had
predicted. Every month one or more papers on this subject are put
in the quant-ph electronic archive or appear in other media. Typical
questions that are discussed are how to achieve optimal, if not perfect
cloning \cite{optimal}, a question which led to the important notion
of quantum disentanglement \cite{terno}; the no-broadcasting theorem
\cite{broadcast}, which is a generalization of no-cloning to general
impure density matrices; $M\to N$ probabilistic cloning \cite{duan,m2n};
rigorous and stronger theorems \cite{lindblad,koashi,jozsa}; and in
particular the relation of optimal cloning to the impossibility of
signaling \cite{gisin,ghosh,dagmar} and bounds on state estimation that
are imposed by the no signaling condition \cite{erika}. The problem
was finally laid to rest by a fundamental result: completely positive
maps, the only ones that can be realized by quantum mechanical systems,
cannot increase distinguishability \cite{graaf}.

There is no doubt that the term ``superluminal'' was one of the causes
this subject became so attractive. Who would not be happy to beat the
relativistic limit on the speed of transmission of information?
Actually, superluminal group velocities have been observed in barrier
tunneling in condensed matter \cite{chiao1,chiao2}. However, special
relativity does not forbid the {\it group\/} velocity to exceed~$c$. It
is the {\it front\/} velocity of a wave packet that is the relevant
criterion for signal transmission, and the front velocity never
exceeds~$c$.

Clearly, the term superluminal has no room in the present
considerations.  The speed of light never enters in them, nor does
the theory of relativity. The issue simply is: given two observers
with an entangled bipartite quantum system (or even numerous such
systems: they are equivalent to a single, bigger, physical system)
it is impossible, by means of local quantum operations (LQO), to
transmit any information whatsoever from one observer to the other,
without transmitting real material objects between them.

This is what quantum mechanics says \cite{interv2}, and a speculative
question could be: what kinds of modifications of quantum mechanics
would allow information transfer by means of LQOs. In the following,
I'll show that a nonlinear evolution of the density matrix $\rho$ would
give such a result. A similar conclusion was also reached by Svetlichny
\cite{svet}, who however only considered the special case of an initial
maximally entangled pure state, and made further restrictive assumptions
(and also used the term ``superluminal'' without ever using the theory
of relativity).

Consider our two familiar observers, Alice and Bob, who have a bipartite
quantum system in a known state $\rho$.  They perform LQOs that are
mathematically represented by positive operator valued measures (POVMs)
with elements $\sum A_j=\1$ and $\sum B_\mu=\1$ respectively. The
probability for the joint result $j\mu$ is

\beq P_{j\mu}=\Tr_A\Tr_B\,(\rho\,A_j\0B_\mu), \eeq
where the double trace is taken on the indices of Alice and of Bob. If
Bob is not informed that Alice got result $j$, the probability that
he gets $\mu$ is

\beq \sum_j P_{j\mu}=\Tr(\rho_B B_\mu), \eeq
where $\rho_B=\Tr_A(\rho)$ is Bob's reduced density matrix. This
result does not depend on Alice's choice of a POVM. This is what
quantum mechanics says, and all this is well known.

If Bob is informed that Alice got result $j$, it follows from (1) that 
the probability that Bob has result $\mu$ is $\Tr_B(\rho_j B_\mu)$, 
where

\beq \rho_j=\Tr_A(\rho A_j)/p_j. \label{rhoj} \eeq
Here the denominator $p_j=\sum_\nu P_{j\nu}$ is the probability that
Alice gets result $j$, so that $\rho_j$ has unit trace as it should.

Therefore everything happens as if the state of Bob's subsystem
actually was $\rho_j$. Quantum mechanics does not claim that this
is true, but also gives no way of showing that this realistic point
of view is false.  It's just a matter of belief. Bernard d'Espagnat
calls Bob's reduced density matrix an ``improper'' mixture
\cite{despagnat}. I also
considered such situations in \cite{compounds}, where I distinguished
pure states, mixtures, and ``compounds'' (a compound is a mixture that
has a unique decomposition into pure states if additional information
is supplied). Here, I am slightly more general: Bob's reduced density
matrix can be split in a unique way into other density matrices (not
only pure states) if Alice reveals to him which result she got.

The question is whether Bob can do that without Alice's help, and
thereby find out which POVM she chose to perform. For this, one has to
violate quantum mechanics in some way, such as cloning, as Nick Herbert
originally proposed. Here, it is essential to assume that cloning
an improper mixture (a ``compound'') means to clone each component
separately, as if the state really was one of these components, with
probability $p_j$, and we merely ignored which one.

Let us suppose that Bob can perform a trace-preserving nonlinear
transformation 

\beq \rho_j\to\tilde{\rho}_j, \eeq
on the unknown ``true'' states of his subsystem. Then his reduced
density matrix evolves as 

\beq \rho_B=\sum_j p_j\rho_j\to\sum_j p_j\tilde{\rho}_j. \eeq
Note that the result is not determined by $\rho_B$ alone, but the
decomposition of $\rho_B$ into a definite set of $\rho_j$ is
essential. In particular, the result depends on the choice made by
Alice of a particular POVM, as Eq.~(\ref{rhoj}) explicitly shows.

If Alice chooses a different POVM, $\sum A_s=\1$, giving Bob states
$=\Tr_A(\rho A_s)/p_s$ with probability $p_s$, Bob would obtain, after
the hypothetical nonlinear evolution,

\beq \rho_B=\sum_s p_s\rho_s\to\sum_s p_s\tilde{\rho}_s. \eeq
Let us call the right hand sides of the last two equations $\rho'$ and
$\rho''$, respectively. In the generic case, they are not equal to each
other and they can be distinguished statistically. If enough copies are
supplied to Bob, he would be able to know which POVM Alice chose.

Needless to say, the assumption of a nonlinear evolution of $\rho$
violates quantum mechanics. Contrarywise, if $\rho$ evolves linearly,
then $\rho'=\rho''$ and no information can be transmitted in this way.
Note that all the discussion is about the density matrix $\rho$. It
is quite possible to have a nonlinear evolution of pure states which
corresponds to a linear evolution of $\rho$. For example, the evolution

\beq {\alpha\choose\beta}\to{1\choose0} \eeq
is nonlinear in terms of pure states, because

\beq {\alpha\choose\beta}\equiv\alpha{1\choose0}+\beta{0\choose1}, \eeq
and the right hand side would give $(\alpha+\beta){1\choose0}$ if the
evolution were linear. 

On the other hand, the same evolution expressed in terms of density
matrices appears as

\beq {{a\qquad b}\choose{\quad
  b^*\quad1-a}}\to{{1\qquad0}\choose{0\qquad0}}. \eeq
This is a linear operation, generated by a pair of Kraus matrices 

\beq {{1\qquad0}\choose{0\qquad0}} \qquad{\rm and} \qquad
  {{0\qquad1}\choose{0\qquad0}}. \eeq

\bigskip In summary,
Nick Herbert's erroneous paper was a spark that generated immense
progress. There also are many wrong papers that have been published
in reputable journals, some of them by renowned scientists. Their bad
influence may last for years. For these, I decline all responsibility. I
was not the referee of these papers and I could not protect the good
reputation of their authors.

\bigskip I am grateful to Erika Andersson, Gilles Brassard, Dagmar
Bru\ss, Chris Fuchs, GianCarlo Ghirardi, Nobu Imoto, Danny Terno,
and Bill Wootters for many clarifying comments and for helping me to
locate some of the references below. This work was supported by the
Gerard Swope Fund and the Fund for Encouragement of Research.

\end{document}